\documentclass[final,3p,times,twocolumn]{elsarticle}

\usepackage{caption}
\usepackage{amssymb}
\usepackage{gensymb}

\usepackage{tabularray}
\usepackage{amsmath}
\usepackage[outline]{contour}
\usepackage{longtable}

\usepackage{tabularx}

\usepackage{multirow}

\journal{Computational Materials Science}

\begin{document}

\begin{frontmatter}



\title{First-Principles Insights into the Site Occupancy of Ta-Fe-Al C14 Laves Phases}


\author[label1]{Nisa Ulumuddin\corref{cor1}} 

\affiliation[label1]{organization={Institute of Physical Metallurgy and Materials Physics, RWTH Aachen University, 52056 Aachen, Germany}}

\author[label1]{Sandra Korte-Kerzel} 

\author[label1]{Zhuocheng Xie\corref{cor2}} 

\cortext[cor1]{Corresponding author \\
Email address: ulumuddin@imm.rwth-aachen.de}
\cortext[cor2]{Corresponding author \\
Email address: xie@imm.rwth-aachen.de}

\begin{abstract}
This study investigates the site occupancy preferences of Al in Ta(Fe$_{1-x}$Al$_x$)$_2$ Laves phases using first-principles calculations, covering Al concentrations from 0 to 50 at.\%. Al atoms exhibit a strong preference for $2a$ Wyckoff sites, with configurations becoming more energetically favorable as these sites reach full occupancy at high Al concentrations. Magnetic configurations were explored, revealing that anti-ferromagnetic ordering is the most favorable at ground states. A metastable defect phase diagram based on the chemical potential of Al was constructed to map site occupancy preferences, where Ta$_4$Fe$_6$Al$_2$ and Ta$_4$Fe$_2$Al$_6$ exhibit the widest chemical potential windows. The correlation between lattice distortions and site occupancy was examined, demonstrating that symmetric Al distributions enhance structural preference.  These findings offer insights into the structural motifs of the Ta-Fe-Al system, providing a foundation for future investigations on structure-property relationships.
\end{abstract}


\end{frontmatter}



\section{Introduction}
\label{intro}

Laves phases exhibit a unique combination of high-temperature stability, oxidation resistance, and hardness, making them attractive for advanced structural and functional applications \cite{livingston1992laves,stein2021laves,xu2021tailoring,sui2021laves}.  They are increasingly utilized in hydrogen storage, superconductors, and wear- and corrosion-resistant materials, while rare-earth-containing Laves phases are particularly valued for their magnetic properties \cite{stein2021laves, yartys2022laves}. However, their inherent brittleness at room temperature—stemming from the dense atomic packing of large (A) and small (B) atoms in AB$_2$ crystal structures—presents a significant challenge for wide applications \cite{laves1934kristallstruktur,laves1935kristallstruktur}. A deeper understanding of their mechanical behavior of Laves phases, including elastic anisotropy, dislocation-mediated plasticity, and crack nucleation and propagation, is essential for improving their structural integrity and unlocking superior functional properties.

\begin{figure*}[!ht]
    \centering
    \includegraphics[width=0.9\linewidth]{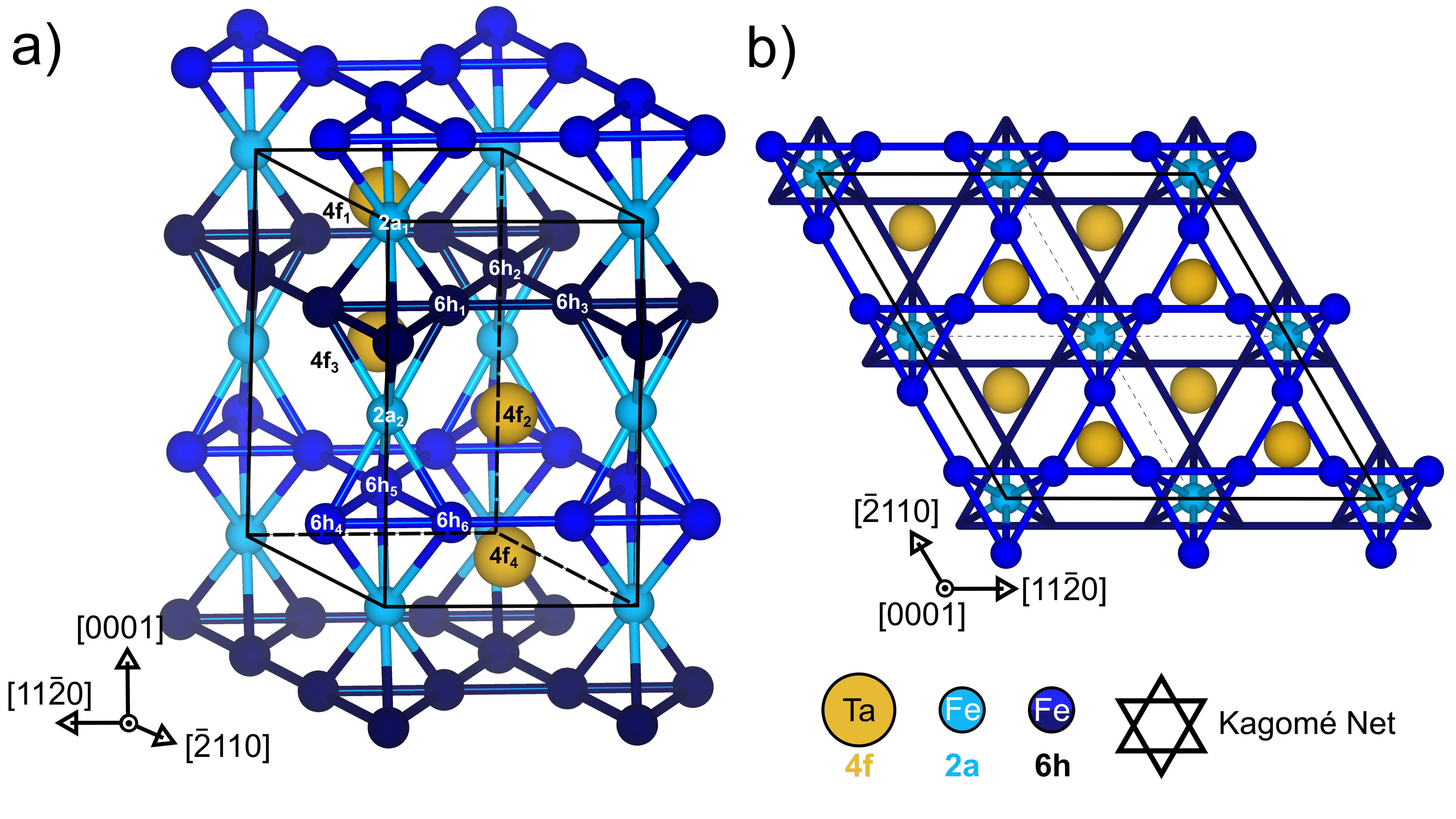}
    \caption{The conventional unit cell of C14 TaFe$_2$ viewed from (a) an off-axis perspective and (b) along the $\langle c \rangle$ axis. The structure consists of alternating triple layers and Kagomé layers along the $\langle c \rangle$ direction. The triple layer consists of the $4f$ Ta atoms (brown) that sandwich the $2a$ Fe atoms (light blue). The Kagomé Fe layers, shown in two different shades of dark blue, are offset, as the atoms in the upper and lower layers do not align on top of one another.}
    \label{fig:1}
\end{figure*}

The C14 Laves phase (space group $P6_{3}/mmc$ and prototype MgZn$_2$) features a hexagonal cell with 12 atoms, composed of alternating stacks of triple layers and Kagomé layers along the $\langle c \rangle$ axis. The A atoms occupy the $4f$ Wyckoff site with a coordination number of 16 (Z=16), sandwiching the smaller B atoms at the $2a$ sites within the triple layers when viewed along the $\langle a \rangle$ direction. The Kagomé layer consists of B atoms at the $6h$ Wyckoff sites (Z=12), forming a sublattice of tetrahedra along the $\langle c \rangle$ axis. These tetrahedra alternate between face-sharing and vertex-sharing, creating trigonal bipyramids joined at their apexes (Figure \ref{fig:1}a). The A atoms are centered within large polyhedral networks of B atoms. Half of the triangular B-atom nets in the Kagomé layer are "capped" by a B-atom in the $2a$ position, while the other half remain "uncapped" when viewed along the $\langle c \rangle$ axis (see Figure \ref{fig:1}b). 

Chemical composition has long been recognized as a key factor in modulating the mechanical properties of Laves phases \cite{takasugi1996deformability,thoma1997elastic,yoshida1997alloying,chen1998factors,takata2016nanoindentation,FREUND2024120124} and their composites \cite{yurchenko2016effect,zubair2019role,zubair2023laves,han2023effects}. Ternary Laves phases hold significant potential for advancing high-entropy alloys given their prevalence as intermetallic phases in these materials \cite{tsai2019intermetallic}. Introducing additional alloying elements to Laves phases can lead to either hardening or softening, depending on the alloying system \cite{takasugi1996deformability,thoma1997elastic,yoshida1997alloying,chen1998factors,takata2016nanoindentation,FREUND2024120124}. However, much of these efforts rely on trial and error, as a fundamental understanding of the mechanical response, particularly in terms of dislocation mechanisms under various stimuli, remains lacking. Designing Laves phases with tailored mechanical properties requires consideration of deviations from the ideal AB$_2$ stoichiometry, which can introduce structural point defects, such as anti-site atoms, vacancies, and ordered or disordered integration of ternary elements. A critical first step in this process is understanding the site occupancy preferences in alloyed Laves phases. 

The site occupancy preferences in ternary C14 Laves phases have been investigated using X-ray diffraction and ab-initio calculations. At the stoichiometric A-composition (33.3 at.$\%$), it is widely agreed upon that the ternary elements predominantly occupy either the $2a$ or $6h$ B sites \cite{majzoub2003rietveld,kerkau2009site,yan2009laves,yamagata2023site}, even when their atomic radius is close to that of the A-atom \cite{yamagata2023site}. The preference for either the $6h$ or the $2a$ sites tends to vary depending on the material system. Some systems \cite{grytsiv2006atom,yan2008ternary,kerkau2009site} show distinct composition-dependent site occupancy preferences, while in others, the occupation of the $2a$ and $6h$ sites is considered random \cite{majzoub2003rietveld,yan2009laves,yan2022site,yamagata2023site}. The stability of the ternary Laves phase has been discussed in terms of the geometric factors, such as the lattice parameters ($c/a$) or atomic radii (r$_B$/r$_A$) ratios \cite{yamagata2023site}. However, the qualitative differences in the progression of the $c/a$ ratio with composition across stoichiometric and off-stoichiometric Laves phases suggest that these simple geometric factors alone are insufficient to predict stability.  For instance, variations in site occupation trends observed in Zr(V$_{1-x}$Co$_x$)$_2$ and Nb(Cr$_{1-x}$Co$_x$)$_2$ \cite{kerkau2009site} indicate that bonding effects play a significant role in determining site occupancy, making each ternary Laves phase system potentially unique.  

In this work, the site occupancy of a common Laves phase that possesses a large solubility range, namely the C14 Ta(Fe,Al)$_2$, was investigated. It has been identified as one of the less brittle Laves precipitates that can form in the $\alpha$-Fe matrix \cite{jones1972laves}. The Fe-Ta-Al phase diagram has been explored extensively by Witusiewicz et al. \cite{witusiewicz2013experimental} and elaborated further by Raghavan et al. \cite{raghavan2013phase}, showing that the homogeneous ternary Ta-Fe-Al C14 Laves phase constitutes a wide range of Fe and Al concentrations, but a narrow range of Ta concentrations. As the solubility of Al in the Ta-Fe-Al Laves phase can reach up to 52 at.$\%$ \cite{witusiewicz2013experimental}, suggesting the mechanical properties of the TaFe$_2$ phase could be tunable through Al substitution. This work explored the site occupancy preferences of Al in Ta(Fe$_{1-x}$Al$_x$)$_2$ Laves phases, with Al concentrations ranging from 0 to 50 at.$\%$ using density functional theory (DFT). The preferable magnetic configurations of each Ta(Fe$_{1-x}$Al$_x$)$_2$ composition were identified. A metastable defect phase diagram of the Ta(Fe$_{1-x}$Al$_x$)$_2$ C14 Laves phases as a function of Al chemical potential was constructed, informed by experimental phase boundaries. Correlations between lattice distortion and site occupancy preference were also examined.

\section{Computational Methods}
\label{methods}

\begin{table}[t]
\centering
\resizebox{\columnwidth}{!}{
\begin{tabular}{cc}
\hline
\textbf{Al-rich}  & \textbf{Al-poor}  \\ 
\textnormal{\textbf{TaAl$_3$ + Fe$_4$Al$_{13}$ + C14}}   & \textnormal{\textbf{C14 + Fe$_\textnormal{BCC}$}}        \\  \hline \hline 
$ \mu_{\textnormal{Ta}} + 3 \mu_{\textnormal{Al}} < E^0_{\textnormal{TaAl}_3}$   &   $ 63 \mu_{\textnormal{Fe}} + 1 \mu_{\textnormal{Al}} + 32 \mu_{\textnormal{Ta}} < E^0_{\textnormal{Fe}_{63}\textnormal{Al}_1\textnormal{Ta}_{32}} $               \\
$ 2 \mu_{\textnormal{Fe}} + 6 \mu_{\textnormal{Al}} + 4 \mu_{\textnormal{Ta}} < E^0_{\textnormal{Fe}_2\textnormal{Al}_6\textnormal{Ta}_4}$  & $ \mu_{\textnormal{Fe}} < E_{\textnormal{Fe}_{\textnormal{BCC}}}^0$                                       \\
$ 4 \mu_{\textnormal{Fe}} + 13 \mu_{\textnormal{Al}} < E^0_{\textnormal{Fe}_4\textnormal{Al}_{13}} $      &   $\mu_{\textnormal{Ta}}^{\textnormal{Al-poor}} = \mu_{\textnormal{Ta}}^{\textnormal{Al-rich}} $        \\ \hline
\end{tabular}
}
\caption{Phase boundaries of the ternary Ta-Fe-Al C14 Laves phase based on the semi-empirical phase diagram \cite{witusiewicz2013experimental}. The solubility limit of Al is 52 at.$\%$ corresponds to Fe$_2$Al$_6$Ta$_4$ at the Al-rich limit. For the Al-poor limit, the bulk energy of Ta$_{32}$Fe$_{63}$Al$_1$ was used. Bulk energies are denoted with a superscript "0" ($E^0$). The chemical potentials $\mu$ of Al, Fe, and Ta lie between these limits. Other phases are assumed to exist in their binary form.}
\label{tab:0}
\end{table}

DFT calculations were performed using the Vienna Ab Initio Simulation Package (VASP) \cite{kresse1996efficient,kresse1993ab}. The Projector Augmented Wave (PAW) potential was employed to construct the planewave basis set \cite{kresse1999ultrasoft}, with the exchange-correlation functional using the generalized gradient approximation (GGA) of Perdew–Burke–Ernzerhof (PBE)  \cite{perdew1996generalized}.  The kinetic energy cutoff for the planewave basis set was set to 550 eV. These basis sets implement frozen core electrons, as bonding is assumed to occur only among valence electrons. In the PAW potentials used, Fe $4s^1, 3d^7$, Ta $6s^1, 5d^4, 5p^6$ and Al $3p^1, 3s^2$ were set as the valence orbitals.
In each geometry optimization calculation, the Fermi level was smeared using the 1$^{st}$ order Methfessel-Paxton method with a $\sigma$ level of 0.1 eV \cite{methfessel1989high}. Spin polarization was enabled in all calculations. The self-consistent field (SCF) cycles and geometric optimization were considered converged when the energy difference reached 10$^{-6}$ eV and the forces were less than 0.02 eV/$\mathrm{\AA{}}$. The Brillouin zone was sampled using the Monkhorst-Pack $k$-point scheme \cite{monkhorst1976special}. Geometry optimizations of the conventional unit cell were performed using a gamma-centered $k$-point grid of 10 $\times$ 10 $\times$ 5. VESTA \cite{momma2011vesta} and OVITO \cite{stukowski2009visualization} were used for visualization.

The C14 Laves phase, possessing a hexagonal crystal structure, features independent lattice parameters $a$ and $c$, and the angles $\alpha, \beta,$ and $\gamma$ are fixed at 90$^{\circ}$, 90$^{\circ}$, and 120$^{\circ}$, respectively, in its conventional unit cell. The ground-state lattice parameters were achieved by fully relaxing the atomic position, cell shape, and cell size. Magnetism was included in the configurational sampling by permuting the spin-up and spin-down arrangement across all Fe atoms, with each configuration considered as a separate structure. Structures with ternary compositions were optimized using the same methodology. This results in the full-cell relaxation of three unique configurations in TaFe$_2$, and many more for ternary compositions, as detailed in Table \ref{tab:1}.   

The site occupancy preferences of Al in Ta(Fe$_{1-x}$Al$_x$)$_2$ Laves phases, where the Ta concentration is held at the stoichiometric composition of 1/3, were explored using DFT. The Al concentration was examined across a range defined by the solubility limit identified by Witusiewicz et al. \cite{witusiewicz2013experimental}, i.e., $ 0 \leq x_\textnormal{Al} \leq 0.52$. Site occupation was explored within one conventional C14 unit cell, containing four formula units of AB$_2$, corresponding to six $6h$ sites, two $2a$ sites, and four $4f$ sites. Al substitution was permuted among the B-sites ($6h$ and $2a$ sites), as previous studies \cite{yamagata2023site,prymak2008phase,kerkau2009site} suggest that A-site substitution is highly unlikely in A-stoichiometric C14 Laves phases, even in cases where the atomic radius of the ternary component is closer to that of the A atom.

The free energy of a structure can be generally defined by Equation \ref{eq:1a}.

\begin{multline}
    F(V,T) =E_{tot}(V) +F_{electron}(V, T) + F_{vibr}(V,T) \\ +F_{magnet}(V,T) – TS_{config}(V).
    \label{eq:1a}
\end{multline}

where E$_{tot}$(V) is the DFT energy of the system,  F$_{electron}$(V, T), F$_{vibr}$(V,T), and F$_{magnet}$(V,T) are the electronic, vibrational and magnetic free energy contributions, respectively, and TS$_{config}$(V) is the configurational entropy term. The energy contributions of F$_{electron}$(V, T), F$_{magnet}$(V,T) and TS$_{config}$(V) in the C14 Laves phase tend to be extremely small relative to F$_{vibr}$(V,T), particularly near room temperature \cite{vasilyev2023comparison,vasilyev2021new}. As displayed in Figure \ref{fig:app1}, the vibrational energy for temperatures less than 500 K accounts for less than 1 \% of the total free energy in Ta(Fe$_{1-x}$Al$_x$)$_2$. Additionally, thermal expansion of lattices at approximately room temperature is typically not significant \cite{vasilyev2023comparison,mayer2003ab, samanta2022experimental}. Therefore, for the remainder of this study, we will use the DFT energy, E$_{tot}$(V), to assess site occupancy preference. 

To determine the most favorable Al site occupancy pattern for each composition, the formation energy of each structure was calculated with respect to the elemental chemical potential of its constituents, as shown in Equation \ref{eq:1}:

\begin{equation}
    \Delta E_F = E^0_{\textnormal{Ta}_x\textnormal{Fe}_{y}\textnormal{Al}_z} -  xE^0_{\textnormal{Ta}} - yE_{\textnormal{Fe}}^0 - z E_{\textnormal{Al}}^0,
    \label{eq:1}
\end{equation}

where E$^0_{\textnormal{Ta}_x\textnormal{Fe}_{y}\textnormal{Al}_z}$ , E$^0_{\textnormal{Ta}}$, E$_{\textnormal{Fe}}^0$ and E$_{\textnormal{Al}}^0$ are the per-atom bulk energies of the Al-occupied structure, BCC Ta, FCC Fe and FCC Al at ground-state, respectively. The values of $x$, $y$, and $z$ represent the percentages of Ta, Fe, and Al in the ternary alloy, respectively.

The thermodynamical stability of the lowest-energy defect structures was determined by computing the convex hull of formation energies for all known stable phases in the ternary Ta-Fe-Al system \cite{witusiewicz2013experimental} and the lowest-energy Ta(Fe$_{1-x}$Al$_x$)$_2$ ground-state structures. The lattice parameters and, consequently, the energy of each phase were optimized using the same calculation parameters. The convex hull was visualized as a ternary diagram using the pymatgen package \cite{ong2008li,ong2010thermal}. A compound is considered stable at 0 K if its energy lies on the convex hull.

To capture the site occupancy trends within TaFe$_{2}$, we used the concept of metastable defect phase diagrams \cite{korte2022defect}, where defect formation energies were computed relative to the pristine TaFe$_{2}$, treating the system as an open thermodynamic environment in equilibrium with reservoirs of its elemental components. This approach is well-suited for studying how Al incorporates into the Fe sublattice within the C14 Laves phase and aligns with previous works on planar defects in Laves phases \cite{tehranchi2024metastable}. The defect formation energy is consequently formulated as Equation \ref{eq:2}:

\begin{equation}
    \Delta E^{defect}_F = E^0_{\textnormal{Ta}\textnormal{Fe}_{2-x}\textnormal{Al}_x} - \left(E^0_{\textnormal{Ta}\textnormal{Fe}_2}  - x\mu_{\textnormal{Fe}} + x \mu_{\textnormal{Al}}\right)
    \label{eq:2},
\end{equation}

where $E^0_{\textnormal{Ta}\textnormal{Fe}_{2-x}\textnormal{Al}_x}$ and $E^0_{\textnormal{Ta}\textnormal{Fe}_2}$ are the per-atom bulk energies of the Ta-Fe-Al Laves phase structure and the pristine TaFe$_2$ structure, respectively. $\mu$ denotes the chemical potential of the elements involved in the substitution. To ensure that the defect phase diagram is within the stability range of the ternary Laves phase, the chemical potentials of Fe and Al were constrained by the experimental phase boundaries \cite{witusiewicz2013experimental,raghavan2013phase}. At the Al-rich (Fe-poor) limit, the ternary Laves phase is at the saturation concentration of 52 at.$\%$ Al \cite{witusiewicz2013experimental}, corresponding to Ta$_4$Fe$_2$Al$_6$, which is in equilibrium with TaAl$_3$ and Fe$_4$Al$_{13}$. In the Al-poor (Fe-rich) limit, the ternary Laves phase Ta$_{32}$Fe$_{63}$Al$_1$, where Al-Al solute interaction is limited, is in equilibrium with the BCC Fe phase. The chemical potential of Ta remains unchanged across these limits, as the Ta concentration is fixed. All chemical potential definitions in this work are based on ground-state conditions ($T$=0 K, $P$=0 atm). The phases at the boundary conditions are also present in the ternary convex hull diagram of the Ta-Fe-Al system at the ground-state, making them relevant to our defect phase diagram. Table 1 lists the phase boundaries for the defect phase diagram. Bulk energies are denoted with a superscript "0" ($E^0$). The defect structure which minimizes the defect formation energy is considered thermodynamically metastable in the open system.

\section{Results}
\label{results}

\begin{figure*}[h!]
    \centering
    \includegraphics[width=0.75\linewidth]{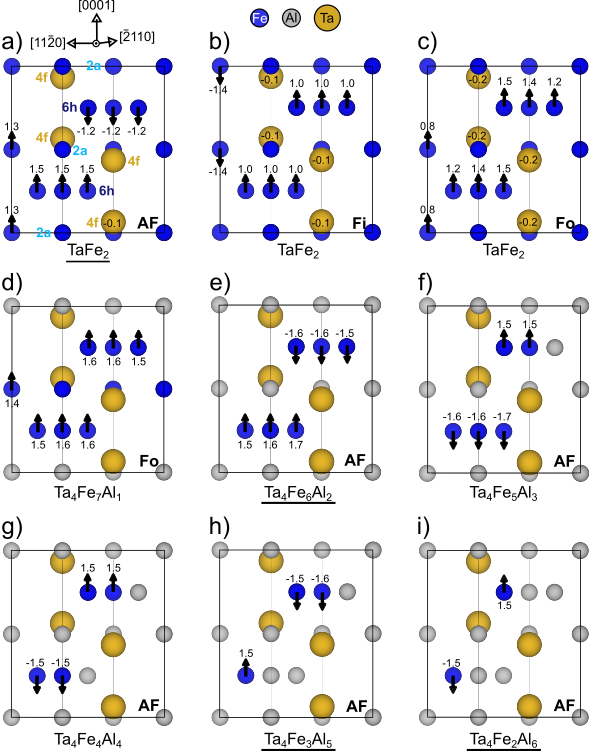}
    \caption{The schematics illustrate (a-c) the possible magnetic configurations in binary TaFe$_2$ and (c-i) the lowest-energy site occupancy orderings and corresponding magnetic configurations in ternary Ta$_4$Fe$_{8-x}$Al$_x$ compositions, as summarized in Table \ref{tab:1}. In (a), the Wyckoff position of each atom is labeled. The magnetic moment of each atom is annotated on or near the atom, with atoms having relatively large magnetic moments labeled according to their spin directions. In the bottom right corner of each schematic, the magnetic configuration is classified as Anti-ferromagnetic (AF) or one of two Ferromagnetic variants (Fo or Fi), based on the ordering between neighboring Kagomé layers. The stable and metastable phases later identified in the phase diagram are underlined.}
    \label{fig:2}
\end{figure*}

\subsection{Site occupancies of C14 Ta(Fe$_{1-x}$Al$_x$)$_2$}

An exhaustive list of all binary and ternary site occupation configurations at 33.3 at.\% Ta, including atomic distribution and magnetic ordering, is given in Table \ref{tab:1} in Appendix. Both atomic and magnetic configurations are given in the same order as the atoms indexed in Figure \ref{fig:1}. As the Al content increases, the number of possible orderings also rises, peaking when the Fe:Al ratio approaches 1:1.  The lowest-energy structures for all compositions, highlighted in bold in Table \ref{tab:1}, are illustrated in Figure $\ref{fig:2}$.
Across all compositions, Al consistently shows a strong preference for occupying the $2a$ sites, as indicated by the lowest-energy configurations (highlighted in bold in Table \ref{tab:1}), which contain the maximum number of $2a$-site substitution for each composition. For instance, in the Ta$_4$Fe$_7$Al$_1$ composition, Al substitution at a $6h$-site ($^{2a}$(Fe, Fe)$^{6h}$(Al Fe Fe, Fe Fe Fe)) is approximately $6\%$ higher in energy compared to the corresponding Al substitution at the $2a$ sites ($^{2a}$(Al, Fe)$^{6h}$(Fe Fe Fe, Fe Fe Fe)), see Table \ref{tab:1}.  In higher Al content compositions, the energy differences suggest that full occupancy of Al at the $2a$ sites is energetically more favorable than partial occupancy. As such, the $2a$ sites are fully occupied by Al in the most favorable configurations. Configurations with partial occupancy of Al at the $2a$ sites are then slightly higher in energy than the lowest-energy configurations. For instance, in the Ta$_4$Fe$_4$Al$_4$ composition, the configuration with partial Al substitution at the $2a$-site ($^{2a}$(Al, Fe)$^{6h}$(Fe Fe Al, Fe Al Al) is only approximately $2\%$ higher in energy compared to the corresponding full Al substitution at the $2a$ sites ($^{2a}$(Al, Al)$^{6h}$(Fe Fe Al, Fe Fe Al)). 

Beyond the strong preference for $2a$ site occupancy, Al solutes at the $6h$ sites tend to distribute evenly between the two Kagomé layers. For compositions with an even number of Al atoms per unit cell, the occupancies of the $6h_{1-3}$ and $6h_{4-6}$ sites in neighboring Kagomé layers are symmetrical in the lowest-energy structures. However, when there is an odd number of Al atoms per unit cell, the $6h$ site occupancies between adjacent Kagomé layers are unequal, with one Kagomé layer containing more Al atoms than the other. The most energetically unfavorable configurations occur when one Kagomé layer is fully substituted while the adjacent Kagomé layer is completely unsubstituted, resulting in the most uneven distribution of Al between them. In the Ta$_4$Fe$_2$Al$_6$ composition, the configuration with full Al substitution at both the $2a$-sites and one of the Kagomé layers ($^{2a}$(Al, Al)$^{6h}$(Fe Fe Al, Al Al Al)) is the least favorable, with an energy approximately $16\%$ higher than the configuration with an even distribution of Al across the two Kagomé layers ($^{2a}$(Al, Al)$^{6h}$(Fe Al Al, Fe Al Al)), see Table \ref{tab:1}. Interestingly, only higher-energy structures (with $\Delta E^{form}_{min} \sim 10-30 \%$) display that partial filling of the $2a$ sites can be more favorable than the full occupancy. For example, in Ta$_4$Fe$_2$Al$_6$,  $^{2a}$(Al, Al) $^{6h}$(Fe Fe Al, Al Al Al) is higher in energy than $^{2a}$(Al, Fe)$^{6h}$(Fe Al Al, Al Al Al).

\subsection{Magnetic configurations of C14 Ta(Fe$_{1-x}$Al$_x$)$_2$}

Among the calculated binary and ternary C14 Ta(Fe$_{1-x}$Al$_x$)$_2$ compositions, anti-ferromagnetic ordering is generally the most energetically favorable configuration, except for Ta$_4$Fe$_7$Al$_1$, where ferromagnetic ordering is more favorable (see Figure \ref{fig:2} and Table \ref{tab:1}). In TaFe$_2$, three low-energy configurations were identified: anti-ferromagnetically ordered Kagomé Fe atoms (AF), and two variations of ferromagnetically ordered Kagomé Fe atoms (Fi and Fo), as shown in Figure \ref{fig:2}a, b and c, respectively. Magnetic ordering has a less significant impact on the formation energy compared to site occupancy in the ternary Laves phases. Across all compositions, the two lowest-energy configurations share the same Al site occupancy but differ in magnetic ordering. The energy differences between these low-energy states with different magnetic configurations are minimal. For example, in the Ta$_4$Fe$_2$Al$_6$ composition, the configuration $^{2a}$(Al, Al)$^{6h}$(Fe Al Al, Fe Al Al) with the ferromagnetic ordering ($^{2a}(0,0) ^{6h}(\uparrow 00,\uparrow 00)$) is only about $0.2\%$ higher in energy than the same configuration with anti-ferromagnetic ordering ($^{2a}(0,0) ^{6h}(\downarrow 00,\uparrow 00)$). 

\subsection{C14 Ta(Fe$_{1-x}$Al$_x$)$_2$ in the ternary phase diagram}

\begin{figure}
    \centering
    \includegraphics[width=1\linewidth]{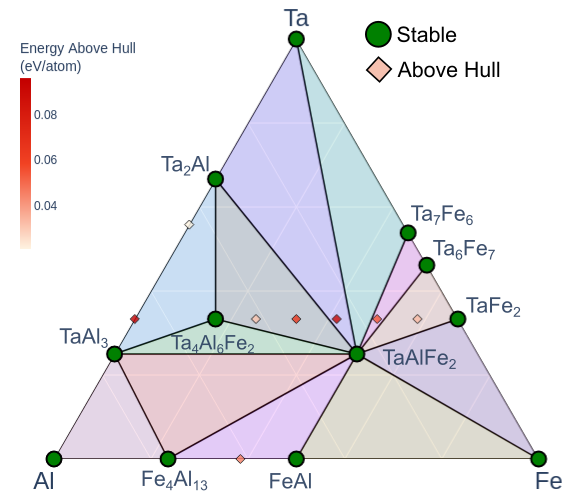}
    \caption{The Ta-Fe-Al ternary phase diagram at 0 K, showing stable (green circle) and unstable (red diamond) phases. The energy above the convex hull (eV/atom) is represented by the color bar. While not all binary Fe-Al stable phases have been included in this phase diagram, they will not contribute further to the stability of Ta(Fe$_{1-x}$Al$_x$)$_2$ phases. The raw data is provided in Table \ref{tab:2} of the Appendix.}
    \label{fig:convexhull}
\end{figure}

Figure \ref{fig:convexhull} depicts the energy differences of the lowest-energy C14 Ta(Fe$_{1-x}$Al$_x$)$_2$ structures relative to the convex hull of the stable phases in the Ta-Fe-Al material system at 0 K. The green circles denote the stable phases while colored diamonds indicate compositions above the hull, thus metastable or unstable. The color gradient corresponds to the energy above the hull, with darker shades indicating higher instability. Among the compositions considered at 33.3 at.\% Ta, the Ta$_4$Fe$_2$Al$_6$ structure appears to be the only stable ternary Ta(Fe$_{1-x}$Al$_x$)$_2$ structure, which corresponds to a 2/3 substitution of the Kagomé $6h$ layers and full substitution of the $2a$-sites. Interestingly, this composition also corresponds to the highest solubility limit of Al in Ta(Fe,Al)$_2$ as reported by Witusiewicz et al. \cite{witusiewicz2013experimental}. The dynamical stability of Ta$_4$Fe$_2$Al$_6$, through the absence of imaginary phonon modes, is shown in Figure \ref{fig:phonons} of the Appendix, implying that it may exist at finite temperatures above 0 K. The Ta$_4$Fe$_2$Al$_6$ phase is thus theoretically accessible though further studies are required to determine its thermal stability at higher temperatures. 

The remaining compositions with negative formation energies are thermodynamically metastable at 0 K and are expected to decompose into more stable phases under equilibrium conditions. 
Among them, the Ta$_4$Fe$_5$Al$_3$ phase has the highest energy above the hull, where the Al site occupancy corresponds to a full substitution of the $2a$ sites but an uneven distribution of Al between adjacent Kagomé $6h$ layers. The degree of instability associated with this uneven Al occupation appears to depend on the element occupying the $2a$ sites, as evidenced by Ta$_4$Fe$_3$Al$_5$, which exhibits a lower energy difference above the hull compared to other compositions. Except for Ta$_4$Fe$_5$Al$_3$, the energies of other Ta(Fe$_{1-x}$Al$_x$)$_2$ compositions remain within 0.07 eV/atom above the hull. The phonon dispersion curves of Ta$_4$Fe$_6$Al$_2$ and  Ta$_4$Fe$_7$Al$_1$, shown in Figure \ref{fig:phonons} of the Appendix, depict dynamical stability. While certain site occupancies are thermodynamically metastable, they may exist under certain chemical potentials or kinetically controlled conditions. The decomposition products will depend on the positioning of the Ta(Fe$_{1-x}$Al$_x$)$_2$ compositions in the phase diagram. 

\section{Discussion}

\begin{figure}
    \centering
    \includegraphics[width=1\linewidth]{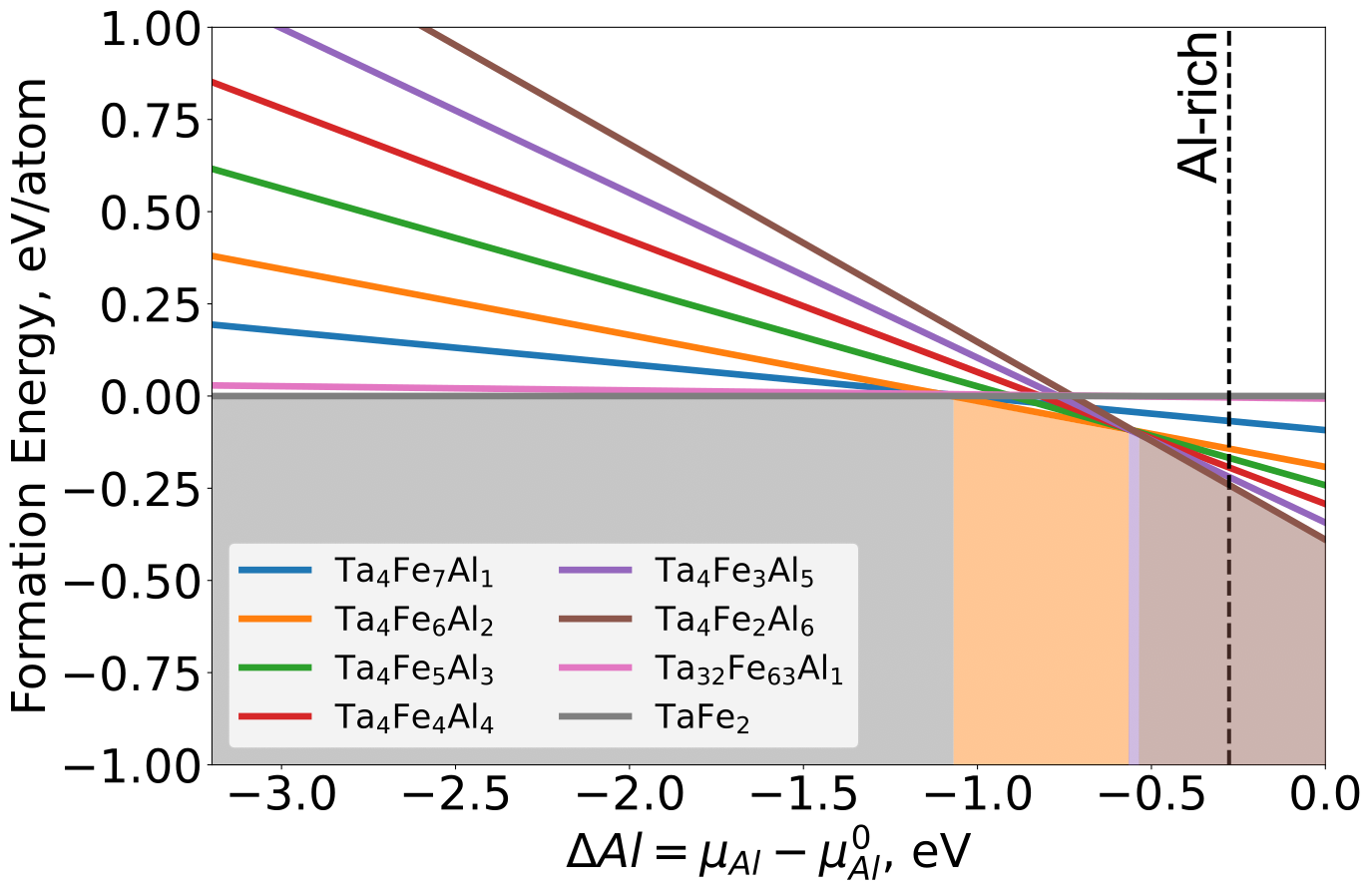}
    \caption{Metastable defect phase diagram of Ta(Fe$_{1-x}$Al$_x$)$_2$ as a function of Al chemical potential, referenced to the bulk energy of Al. The phase boundary, shown as dotted lines in the diagram, defines the chemical potential window of the C14 Laves phase structure. The composition labels refer to the most energetically favorable site occupancy and magnetic moment orderings as displayed in Figure \ref{fig:2}. The chemical potential range in which each structural motif is dominant is highlighted by the colored area beneath the line.}
    \label{fig:phase}
\end{figure}

\begin{figure}[h!]
    \centering
    \includegraphics[width=1\linewidth]{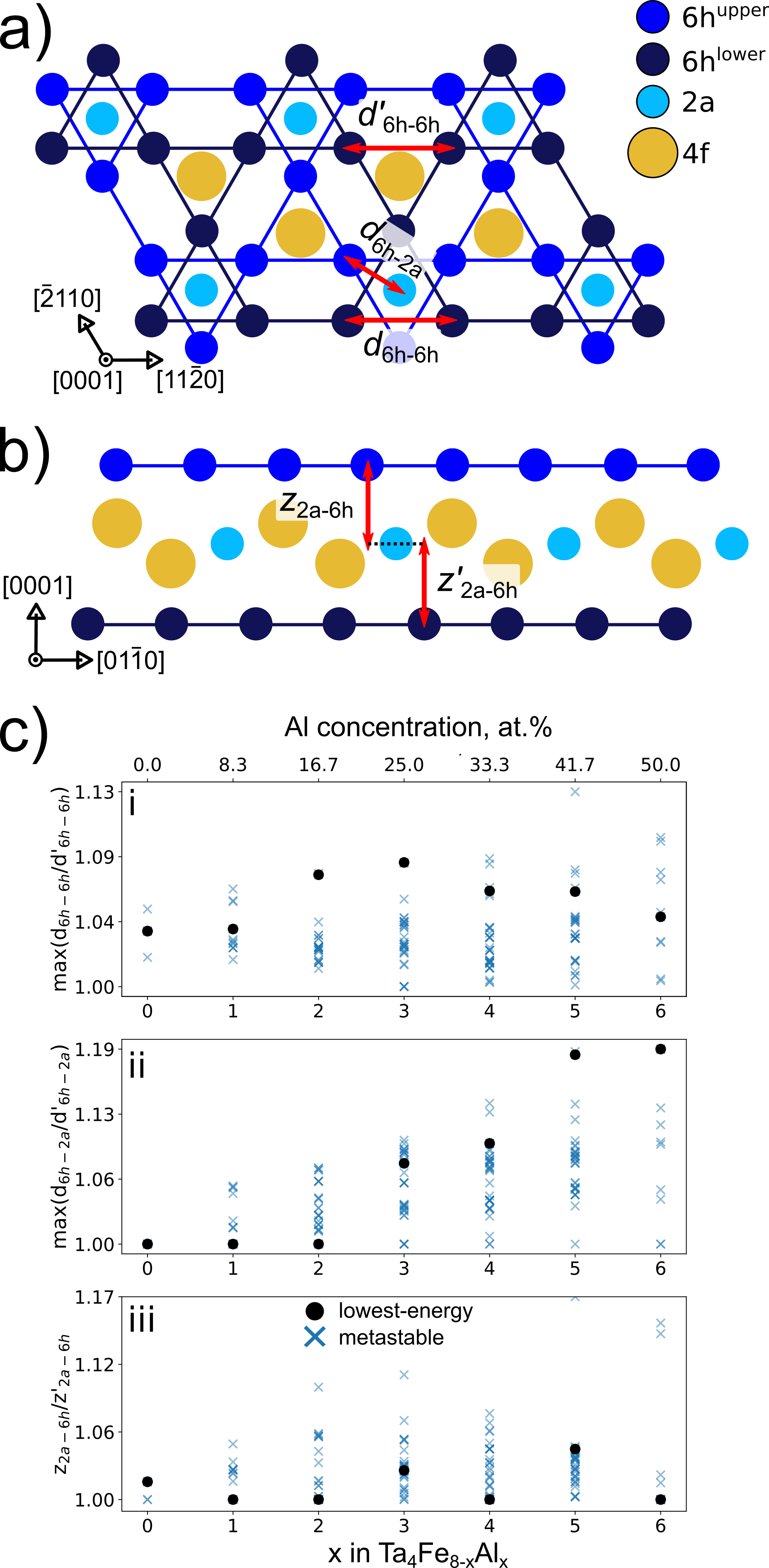}
    \caption{Descriptors of lattice distortion within the C14 Laves phase structure. Schematics show (a) Kagomé layer distortion and (b) interlayer distortion along the $\langle c \rangle$ axis.  The first distortion parameter is the ratio of B-B atom distances in the small capped triangles ($d_{6h-6h}$) and uncapped triangles ($d'_{6h-6h}$). The second parameter is the distortion of the trigonal pyramidal bonds between the Kagomé B atoms (at position $6h$) and the triple-layer B atom (at position $2a$), measured as the ratio of the maximum to the minimum $d_{6h-2a}$. The third parameter is based on the interplanar distance between the Kagomé layer and the $2a$ atom of the triple layer. The ratio is taken between the interplanar distances of the upper ($z_{2a-6h}$) and the lower layers ($z'_{2a-6h}$). The corresponding distortion parameters for each composition are respectively shown in (c) i, ii and iii.}
    \label{fig:distort}
\end{figure}

\subsection{Site occupancy preference in the chemical potential space}

Figure \ref{fig:phase} illustrates how the site occupancy preference patterns evolves with the chemical potential of Al. In this context, each Ta(Fe$_{1-x}$Al$_x$)$_2$ composition is treated as a defect motif within a host matrix, rather than a phase of a closed system.
The chemical potential constraint, based on the experimental phase boundaries as listed in Table \ref{tab:0}, was imposed to ensure the preference of C14 Ta(Fe$_{1-x}$Al$_x$)$_2$ in the chemical potential space. A large region of the chemical potential window is obtained because of the assumption that an extremely low concentration of Al (1 at.\% Al) can be achieved, where solute-solute interactions are limited. Site occupancy patterns with compositions between 0.083 and 1 at.\% Al are expected to show higher preference in the lower chemical potential range, but are not included in this study.   The defect phase diagram illustrates that certain structural motifs dominate larger chemical potential ranges, e.g, Ta$_4$Fe$_6$Al$_2$ and Ta$_4$Fe$_2$Al$_6$, while other ternary compositions, such as Ta$_4$Fe$_7$Al$_1$, remain unfavorable throughout the accessible Al chemical potential range from -6.7 to -0.3 eV. 
The metastability of the Ta$_{32}$Fe$_{63}$Al$_1$ motif relative to TaFe$_2$ in the low Al chemical potential range represents the random solute distribution in the diluted alloy. Within the chemical potential range from -1.1 to -0.6 eV, Ta$_4$Fe$_6$Al$_2$ with full Al substitution at the $2a$-sites and anti-ferromagnetic ordering is the dominant structural motif, despite not being a stable phase in the convex hull diagram (Figure \ref{fig:convexhull}). 
Above -0.6 eV, the Ta$_4$Fe$_2$Al$_6$ configuration with anti-ferromagnetic ordering becomes the dominant defect state. A narrow chemical potential window around -0.6 eV was obtained, where Ta$_4$Fe$_3$Al$_5$ ($^{2a}$(Al, Al)$^{6h}$(Fe Fe Al, Fe Al Al)) becomes the most favorable site occupancy pattern, though with only minor energy differences compared to other configurations. The intersection of formation energy lines for the metastable defect states near -0.6 eV implies that these site occupancies could be transition states before stabilizing in the Ta$_4$Fe$_2$Al$_6$ phase. Kinetically controlled processes may increase the prominence of other intermediate site occupancy patterns within this chemical potential range, especially if dynamical stability is indicated. 
The high energies associated with site occupancies containing relatively high Al substitution at low Al chemical potentials indicate that Al occupancy in the Kagomé layers is less favorable at low Al concentrations. Once the $2a$ sites are fully occupied by Al, additional substitution at the $6h$ sites (within the Kagomé layer) requires a significantly higher chemical potential to serve as the driving force.

\subsection{Site occupancy vs. lattice distortions}

The correlation between site occupancy preference and geometric descriptors offers a practical approach for predicting the stability of ternary Laves phases. In a previous study \cite{yamagata2023site}, geometric factors, such as the lattice parameters ($c/a$) or atomic radii (r$_B$/r$_A$) ratios, were insufficient for predicting the stability of the ternary Laves phase.
In this study, the lattice distortions resulting from the substitution of Al in Ta(Fe$_{1-x}$Al$_x$)$_2$ were characterized using three geometric descriptors, $max(d_{6h-2a}/d^{\prime}_{6h-2a})$, $max(d_{6h-6h}/d^{\prime}_{6h-6h})$, and $z_{2a-6h}/z^{\prime}_{2a-6h}$, as illustrated in Figure \ref{fig:distort}.
Distortions in the Kagomé nets are quantified by the deviation of the $max(d_{6h-6h}/d^{\prime}_{6h-6h})$ and $max(d_{6h-2a}/d^{\prime}_{6h-2a})$ ratios from 1. The $max(d_{6h-6h}/d^{\prime}_{6h-6h})$ ratios show that the triangular nets of the Kagomé layers are not of equal lengths, even in the binary TaFe$_2$, as the triangles capped with $2a$ atoms are slightly longer than the uncapped triangles. This phenomenon has also been seen in other C14 Laves phase structures \cite{johnston1992structure}.

It is reasonable to hypothesize that structures with higher lattice distortion would contain higher internal strain and thus higher energy. However, interestingly, several metastable motifs exhibit much lower lattice distortions across all three ratios than the lowest-energy structures. The metastable structures with the lowest Kagomé distortions $max(d_{6h-6h}/d^{\prime}_{6h-6h})$ contain monoelement Al Kagomé layers, which are the highest in energy and thus the least favorable as previously mentioned.
Structures with $z_{2a-6h}/z^{\prime}_{2a-6h}$ ratios of 1, indicating equidistant interlayer distances perpendicular to the basal plane, correspond to structures with a symmetric Al distribution between adjacent Kagomé layers. The $z_{2a-6h}/z^{\prime}_{2a-6h}$ ratio demonstrates to be a valuable geometric descriptor for predicting the metastability of site occupancy, as it consistently identifies structures with symmetric Al distributions and aligns with the predominant defect states such as Ta$_4$Fe$_6$Al$_2$ and Ta$_4$Fe$_2$Al$_6$.

\subsection{Comparison with experiments}

\begin{figure}[h]
    \centering
    \includegraphics[width=1\linewidth]{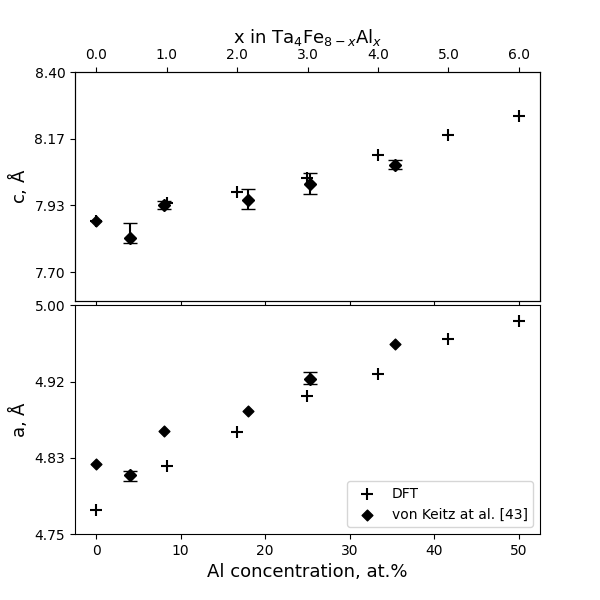}
    \caption{DFT-calculated (+) and experimentally-measured ($\blacklozenge$) \cite{keitz2021laves} $a$ and $c$ lattice constants with increasing Al content in Ta$_4$Fe$_{8-x}$Al$_x$.}
    \label{fig:2a}
\end{figure}

The C14 Ta(Fe$_{2-x}$Al$_x$)$_2$ Laves phases were synthesized and characterized at room temperature as reported by von Keitz et al. \cite{keitz2021laves} and Gasper et al. \cite{gasper2024mechanical}. Figure \ref{fig:2a} displays the $a$ and $c$ lattice constants of the lowest-energy structures predicted by DFT, compared with Ta(Fe$_{1-x}$Al$_x$)$_2$ samples synthesized and measured using X-ray diffraction by von Keitz et al. \cite{keitz2021laves}. The binary TaFe$_2$ converged to lattice parameters of $a$ = 4.784 $\mathrm{\AA{}}$ and $c$ = 7.843 $\mathrm{\AA{}}$. Both lattice constants increase with increasing Al content, consistent with the experimentally observed trend at room temperature, indicating the relevance of the simulated DFT cells to experimental samples. By interpolating between the simulated DFT values, the differences from experimental values are at most approximately 2.7\% for the $c$ constant and 0.4\% for the $a$ constants. Although the difference in the $c$ values is relatively larger, it also has a higher margin of error in von Keitz's experiments \cite{keitz2021laves}. In the study by Gasper et al. \cite{gasper2024mechanical}, Ta concentrations were maintained at 33 \%, while the Fe:Al ratio varied with values of 2, 1, and 0.5, according to energy-dispersive X-ray spectroscopy measurements. These compositions correspond to Ta$_4$Fe$_{5.4}$Al$_{2.7}$,  Ta$_4$Fe$_{4}$Al$_{4}$, and Ta$_4$Fe$_{2.7}$Al$_{5.4}$, which lie between the energetically favorable configurations Ta$_4$Fe$_6$Al$_2$ and Ta$_4$Fe$_2$Al$_6$ identified in Figure \ref{fig:phase}. Von Keitz et al. \cite{keitz2021laves} reported different compositions of Ta$_{4.2}$Fe$_{7.3}$Al$_{0.53}$, Ta$_{4.0}$Fe$_{7.0}$Al$_{1.0}$, Ta$_{3.7}$Fe$_{6.2}$Al$_{2.1}$, Ta$_{3.9}$Fe$_{5.1}$Al$_{3.0}$, and Ta$_{3.7}$Fe$_{4.1}$Al$_{4.2}$, where the Ta concentrations showed minor deviations from 33 \%. The deviations between the experimentally synthesized compositions and the theoretically predicted compositions may arise from two possible aspects. First, the experimentally synthesized compositions could be a mixture of Ta$_4$Fe$_6$Al$_2$ and Ta$_4$Fe$_2$Al$_6$ motifs, possibly due to heterogeneous chemical distributions within the samples. Alternatively, they might contain a mixture of stable and metastable phases, as the energy differences among these phases are relatively small in the high Al chemical potential regime.
Figure \ref{fig:phase} shows that Ta$_4$Fe$_5$Al$_3$, Ta$_4$Fe$_4$Al$_4$, and Ta$_4$Fe$_3$Al$_5$ intersect with the thermodynamically stable configuration Ta$_4$Fe$_2$Al$_6$ at a chemical potential of -0.6 eV relative to bulk Al. This suggests that the experimental Ta-Fe-Al samples containing these structural motifs could correspond to a chemical potential difference of approximately -0.6 eV relative to bulk Al, where these stable and metastable configurations exhibit minor energy differences. Further investigations into these correlations would require high-resolution transmission electron microscopy on these ternary C14 Laves phase samples.

Beyond geometric and compositional considerations, magnetic configurations are crucial in determining the ground-state structure and energy of the Ta-Fe-Al Laves phases. While limited data are available for TaFe$_2$ and its ternary compositions, it is known that the TaFe$_2$ Laves phase is generally considered paramagnetic at room temperature \cite{kai1970magnetic,horie2010magnetic}. However, the exact Néel temperature remains unclear due to the low magnetization observed in stoichiometric TaFe$_2$ \cite{kai1970magnetic}. Yamada et al. \cite{yamada1995magnetic} reported anti-ferromagnetic-like behavior near 10 K, which aligns with the lowest-energy structure among the three magnetic configurations identified in our calculations (Table \ref{tab:1}). 

Table \ref{tab:1} shows several favorable magnetic configurations for each Al site occupancy variant, involving different magnetic coupling combinations between neighboring Kagomé and triple layers. Since Ta is weakly magnetic and Al is non-magnetic, their contributions to the overall magnetism are minimal. Magnetization measurements by Yamada et al. \cite{yamada1995magnetic} suggest that ternary Ta(Fe$_{1-x}$Al$_x$)$_2$ compositions exhibit anti-ferromagnetic behavior, which aligns with our DFT findings. Specifically, the lowest-energy magnetic configurations for all compositions are anti-ferromagnetic, except for Ta$_4$Fe$_7$Al$_1$ which exhibits ferromagnetic behavior (see Figure \ref{fig:2}), which is metastable according to the phase diagram in Figure \ref{fig:convexhull} and does not present itself as a dominant defect state in metastable defect phase diagram (see Figure \ref{fig:phase}), therefore is not expected to be prominent in experimental samples.  

\subsection{Comparison with isostructural systems}

The Ti-Fe-Al system is geometrically similar to Ta-Fe-Al as the difference in metallic radii between Ti and Ta is minimal (approximately 1 pm) \cite{pauling1947atomic}. It has been previously studied by Yan et al. \cite{yan2022site} using X-ray and neutron diffraction and DFT. The ground-state site occupations of Al in the Ta-Fe-Al C14 Laves phase are similar to those calculated in the Ti-Fe-Al system, where Al occupies all the $2a$ sites before increasingly filling the $6h$ sites. This was also reflected in the site occupancy profile obtained through X-ray diffraction, where Al shows a slightly higher preference to occupy $2a$ sites rather than $6h$ sites at low Al concentrations, although both sites are filled up simultaneously \cite{yan2022site}. 

Magnetically, the Ti-Fe-Al can show distinct ordering to Ta-Fe-Al at low temperatures, depending on the Al concentration. Transition from anti-ferromagnetism to ferromagnetism occurs in the Ti-Fe-Al system after a certain Al concentration, which is not observed in Ta-Fe-Al \cite{yan2022site,doi:10.1143/JPSJ.69.225, yamada1995magnetic}. Therefore, while Ti and Ta are both weakly magnetic, they can impact the magnetic coupling between Fe layers differently. A study on Al-substituted NbFe$_2$ reveals that both ferro- and anti-ferromagnetic spin fluctuations coexist in binary and ternary systems \cite{yamada1993nmr}. Ta(Fe$_{1-x}$Al$_x$)$_2$ is stated to be magnetically similar to Nb(Fe$_{1-x}$Al$_x$)$_2$ because of their similar $d$-electron numbers as well as lattice constants \cite{yamada1993nmr}. Our DFT calculations corroborate these observations, showing minor energy differences between ferromagnetic and anti-ferromagnetic ordering among Fe atoms. It is expected that the Ta-Fe-Al system exhibits similar magnetic characteristics, with ferromagnetic and anti-ferromagnetic configurations likely degenerate in energy above the ground-state \cite{yamada1993nmr,horie2010magnetic}. 

\section{Conclusions}
In this work, the site occupancy preferences and magnetic configurations of Ta(Fe$_{1-x}$Al$_x$)$_2$ C14 Laves phases across a wide range of Al concentrations were investigated using DFT. The key outcomes are as follows:

\begin{itemize}
\item Al shows a strong preference for occupying the $2a$ sites across all Ta(Fe$_{1-x}$Al$_x$)$_2$ compositions. Full Al substitution at the $2a$ sites is energetically favorable, while configurations with partial occupancy at the $2a$ sites or Al substitution at the $6h$ sites result in higher energy configurations.

\item The distribution of Al between adjacent Kagomé layers is symmetric in the lowest-energy configurations. Even Al distribution is energetically favorable, while configurations with one fully substituted Kagomé layer and the other unsubstituted, are the least favorable.

\item Ta$_4$Fe$_2$Al$_6$ is the only thermodynamically stable ternary Ta(Fe$_{1-x}$Al$_x$)$_2$ structure at 0 K, corresponding to the highest solubility limit of Al in Ta(Fe,Al)$_2$. Other compositions are thermodynamically metastable and are expected to decompose into more stable phases under equilibrium conditions.

\item A defect phase diagram was constructed as a function of Al chemical potential, revealing that Fe$_6$Al$_2$Ta$_4$ and Fe$_2$Al$_6$Ta$_4$ dominate the largest chemical potential ranges, while other compositions remain less favorable, exhibiting narrow or no chemical potential windows.

\item Anti-ferromagnetic ordering is generally the most favorable magnetic configuration across all compositions. Although magnetic fluctuations and degenerate configurations suggest complex magnetic behavior, the influence of magnetism on formation energy is relatively minor compared to site occupancy.

\item Site occupancy preference correlates with lattice distortion, with symmetry in the interlayer distances ($z_{2a-6h}/z^{\prime}_{2a-6h}$) emerging as a strong geometric descriptor, with minimal distortion correlating with favorable defect configurations in even-numbered Al compositions.

\item The favorable compositions identified theoretically align with experimentally synthesized Ta-Fe-Al Laves phases, though minor deviations in composition suggest heterogeneity in the chemical distribution or free energy contributions influencing the prominence of structural motifs.

\end{itemize}

\section{Declaration of competing interest}
On behalf of all authors, the corresponding author states that there is no conflict of interest.

\section{Data Availability}
Data will be made available on request.

\section{Acknowledgements}
This project has received funding from the European Research Council (ERC) under the European Union’s Horizon 2020 research and innovation programme (grant agreement No. 852096 FunBlocks). The authors gratefully acknowledge the computing time provided to them at the NHR Center NHR4CES at RWTH Aachen University (project number p0020330). This is funded by the Federal Ministry of Education and Research, and the state governments participating on the basis of the resolutions of the GWK for national high performance computing at universities (www.nhr-verein.de/unsere-partner). The data used in this publication was managed using the research data management platform Coscine with storage space granted by the Research Data Storage (RDS) of the DFG and Ministry of Culture and Science of the State of North Rhine-Westphalia (DFG: INST222/1261-1 and MKW: 214-4.06.05.08 - 139057).

\section{Appendix}
\renewcommand{\thefigure}{A\arabic{figure}}
\renewcommand{\thetable}{A\arabic{table}}
\renewcommand{\theequation}{A\arabic{equation}}
\setcounter{table}{0}
\setcounter{figure}{0}
\setcounter{equation}{0}

\begin{figure*}[!hbt]
    \centering
    \includegraphics[width=0.65\linewidth]{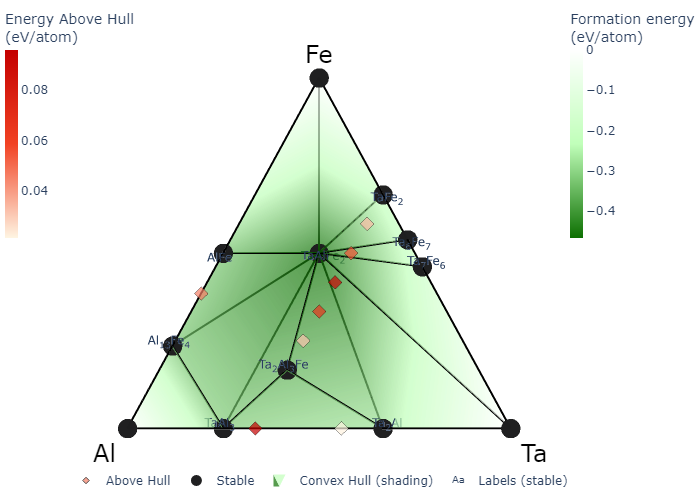}
    \caption{The Ta-Fe-Al convex hull, where the formation energy is given as the shading gradient in the phase space, while energy above the hull is indicated by the (red) shade of the diamond-shaped data points. Stable phases are represented by black circles. Note that not all binary Fe-Al stable phases are included in this diagram. }
    \label{fig:3Dconvexhull}
\end{figure*}

\subsection{Harmonic Approximation of Vibrational Energy}

The Hemholtz free energy is given in Equation \ref{eq:fvib} \cite{togo2023implementation}:

\begin{multline}
    F_{vib} = -k_BT ln Z  \\
    = \phi + \frac{1}{2}\sum_{q\nu} \hbar \omega (\textbf{q} \nu) + k_BT \sum_{\textbf{q}\nu} ln[1-exp(-\hbar \omega(\textbf{q}\nu)/k_BT)],
    \label{eq:fvib}
\end{multline}

where $q$, $\nu$, and $\hbar$ are wave vector, band index, and Planck's constant, respectively. The vibrational frequency $\omega$ was determined using the Harmonic Approximation. Within the finite-displacement supercell approach, atomic displacements are introduced within Ta(Fe,Al)$_2$ supercells of minimum lattice vector dimensions of 10 \AA{}. To ensure accurate force calculations, DFT simulations were performed with a K-points per reciprocal atom (KPPRA) value of 1000. The eigenvalue problem of the dynamical matrix was then solved using the Phonopy package \cite{togo2023implementation}.

\begin{figure}[!ht]
    \centering
    \includegraphics[width=1\linewidth]{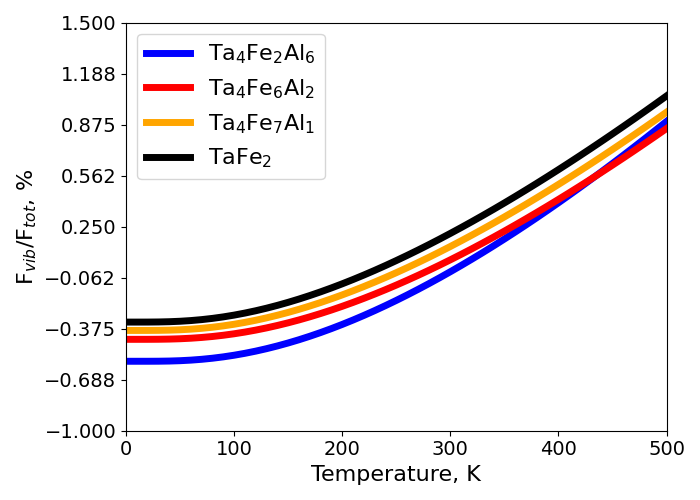}
    \caption{The contribution of vibrational energy towards the total free energy with respect to temperature for TaFe$_2$, Ta$_4$Fe$_7$Al$_1$, Ta$_4$Fe$_6$Al$_2$ and Ta$_4$Fe$_2$Al$_6$.}
    \label{fig:app1}
\end{figure}

\begin{figure}[!ht]
    \centering
    \includegraphics[width=1\linewidth]{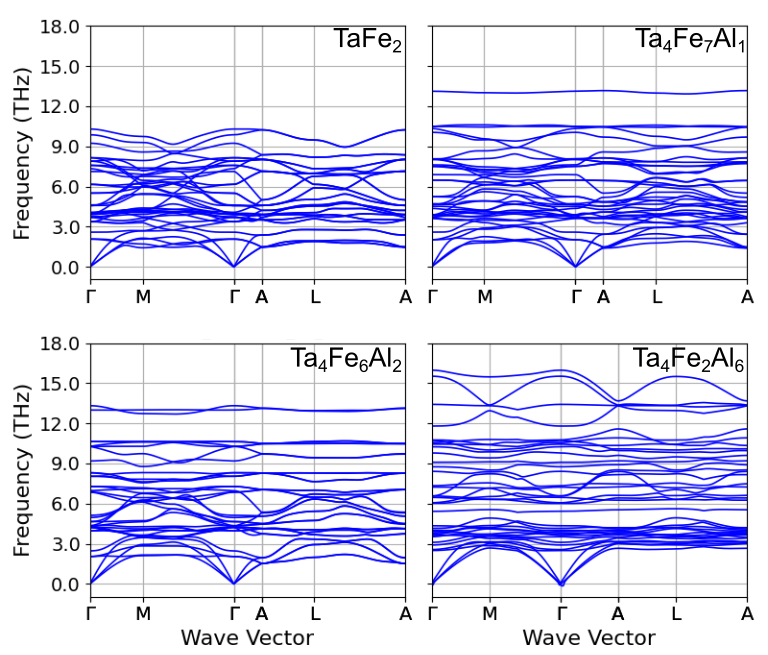}
    \caption{Phonon dispersion of TaFe$_2$, Ta$_4$Fe$_7$Al$_1$, Ta$_4$Fe$_6$Al$_2$ and Ta$_4$Fe$_2$Al$_6$}
    \label{fig:phonons}
\end{figure}

Using only the vibrational energy and electronic (DFT) energy contribution to define the total energy, $ F = E_{DFT} + F_{vib}$,
The fractions of F$_{vib}$ in F of TaFe$_2$, Ta$_4$Fe$_7$Al$_1$, Ta$_4$Fe$_6$Al$_2$ and Ta$_4$Fe$_2$Al$_6$ are given as $100\cdot$F$_{vib}$/F, shown in Figure \ref{fig:app1}, and the corresponding band structures are shown in Figure \ref{fig:phonons}.

\clearpage
\onecolumn 

\begin{longtable}{|l|l|l|l|l|l|}
    \hline
        \textbf{Composition} & 	\textbf{Atomic Ordering} & \textbf{Magnetic Configuration} & \textbf{$\mu_B^{tot}$} & \textbf{E$^\mathrm{form}$} & \textbf{$\Delta$ E$^\mathrm{form}_\mathrm{min}$, \%} \\ \hline
        TaFe$_2$ & \textbf{$^{2a}$(Fe, Fe)$^{6h}$(Fe Fe Fe, Fe Fe Fe)  } & \textbf{$^{2a}(\uparrow, \uparrow)^{6h}(\uparrow \uparrow \uparrow, \downarrow \downarrow \downarrow )$ } & \textbf{ 3.20} & \textbf{-0.21} & \textbf{0} \\ 
        (Ta$_{0.33}$Fe$_{0.67}$) & $^{2a}$(Fe, Fe)$^{6h}$(Fe Fe Fe, Fe Fe Fe) & $^{2a}(\downarrow, \downarrow)^{6h}(\uparrow \uparrow \uparrow, \uparrow \uparrow \uparrow)$ & 1.76 & -0.21 & -1.75 \\ 
        ~ & $^{2a}$(Fe, Fe)$^{6h}$(Fe Fe Fe, Fe Fe Fe) & $^{2a}(\uparrow, \uparrow)^{6h}(\uparrow \uparrow \uparrow, \uparrow \uparrow \uparrow )$ & 8.59 & -0.21 & -2.79 \\ \hline
        Ta$_4$Fe$_7$Al$_1$ & \textbf{$^{2a}$(Al, Fe)$^{6h}$(Fe Fe Fe, Fe Fe Fe)} & \textbf{$^{2a}(0, \uparrow) ^{6h} (\uparrow \uparrow \uparrow, \uparrow \uparrow \uparrow)$ } & \textbf{9.75} & \textbf{-0.27} & \textbf{0} \\ 
        (Ta$_{0.33}$Fe$_{0.58}$Al$_{0.08}$) & $^{2a}$(Al, Fe)$^{6h}$(Fe Fe Fe, Fe Fe Fe) & $^{2a} (0, \uparrow) ^{6h} (\uparrow \uparrow \uparrow, \downarrow \downarrow \downarrow)$ & 1.84 & -0.26 & -0.56 \\ 
        ~ & $^{2a}$(Al, Fe)$^{6h}$(Fe Fe Fe, Fe Fe Fe) & $^{2a} (0, \uparrow) ^{6h} (\downarrow \downarrow \downarrow, \uparrow \uparrow \uparrow)$ & 1.81 & -0.26 & -0.56 \\
        ~ & $^{2a}$(Al, Fe)$^{6h}$(Fe Fe Fe, Fe Fe Fe) & $^{2a} (0, \downarrow) ^{6h} (\uparrow \uparrow \uparrow, \uparrow \uparrow \uparrow) $ & 5.37 & -0.26 & -2.27 \\ 
        ~ & $^{2a}$(Fe, Fe)$^{6h}$(Al Fe Fe, Fe Fe Fe) & $^{2a} (\uparrow, \uparrow ) ^{6h} (\uparrow \uparrow 0, \uparrow \uparrow \uparrow) $ & 9.34 & -0.25 & -6.06 \\ 
        ~ & $^{2a}$(Fe, Fe)$^{6h}$(Al Fe Fe, Fe Fe Fe) & $^{2a}(\uparrow, \uparrow) ^{6h} (\uparrow \uparrow 0, \downarrow \downarrow \downarrow)$ & 1.66 & -0.25 & -7.47 \\ \hline
        Fe$_6$Al$_2$Ta$_4$ & \textbf{$^{2a}$(Al,Al) $^{6h}$(Fe Fe Fe, Fe Fe Fe)} & \textbf{$^{2a}(0, 0)^{6h}(\downarrow \downarrow \downarrow, \uparrow \uparrow \uparrow)$} & \textbf{0.00} & \textbf{-0.32} & \textbf{0} \\ 
        (Ta$_{0.33}$Fe$_{0.5}$Al$_{0.17}$) & $^{2a}$(Al, Al)$^{6h}$(Fe Fe Fe,Fe Fe Fe) & $^{2a}(0, 0)^{6h}(\uparrow \uparrow \uparrow, \uparrow \uparrow \uparrow)$ & 9.03 & -0.32 & -1.24 \\ 
        ~ & $^{2a}$(Fe, Fe)$^{6h}$(Fe Fe Al, Fe Fe Al) & $^{2a}(\uparrow, \uparrow)^{6h} (\uparrow \uparrow 0,\uparrow \uparrow 0)$ & 8.18 & -0.30 & -7.24 \\ 
        ~ & $^{2a}$(Fe, Fe)$^{6h}$(Fe Fe Al, Al Fe Fe) & $^{2a}(\uparrow, \uparrow) ^{6h} (\uparrow \uparrow 0,0 \uparrow \uparrow)$ & 8.41 & -0.30 & -7.26 \\ 
        ~ & $^{2a}$(Al, Fe)$^{6h}$(Fe Fe Al, Fe Fe Fe) & $^{2a}(0, \uparrow)^{6h}(\uparrow \uparrow 0, \uparrow \uparrow \uparrow)$ & 8.13 & -0.29 & -9.34 \\ 
        ~ & $^{2a}$(Al, Fe)$^{6h}$(Fe Fe Al, Fe Fe Fe) & \textbf{$^{2a}(0, \uparrow) ^{6h} (\uparrow \uparrow 0, \downarrow \downarrow \downarrow)$} & -0.57 & -0.29 & -10.33 \\ 
        ~ & $^{2a}$(Fe, Fe)$^{6h}$(Fe Fe Al, Fe Fe Al) & $^{2a} (0,0) ^{6h} (\downarrow \downarrow 0, \uparrow \uparrow 0)$ & 0.01 & -0.29 & -12.18 \\ 
        ~ & $^{2a}$(Fe, Fe)$^{6h}$(Fe Fe Al, Al Fe Fe) & $^{2a}(\uparrow, \uparrow) ^{6h} (\downarrow \downarrow 0, 0 \uparrow \uparrow)$ & 3.29 & -0.28 & -14.28 \\ 
        ~ & $^{2a}$(Fe, Fe)$^{6h}$(Fe Al Al, Fe Fe Fe) & $ ^{2a}( \uparrow, \uparrow) ^{6h} (\uparrow 0 0 , \downarrow \downarrow \downarrow)$ & 7.29 & -0.25 & -21.59 \\ 
        ~ & $^{2a}$(Fe, Fe)$^{6h}$(Fe Al Al, Fe Fe Fe) & $^{2a} (\uparrow, \uparrow) ^{6h} (\uparrow 0 0 , \uparrow \uparrow \uparrow)$ & 0.73 & -0.25 & -22.85 \\ 
        ~ & $^{2a}$(Fe, Fe)$^{6h}$(Fe Al Al, Fe Fe Fe) & $^{2a}(\downarrow, \uparrow) ^{6h} (\uparrow 0 0 , \downarrow \downarrow \downarrow)$ & -2.67 & -0.25 & -23.13 \\ \hline
        Ta$_4$Fe$_5$Al$_3$ & \textbf{$^{2a}$(Al, Al)$^{6h}$(Fe Fe Al, Fe Fe Fe)} & \textbf{$^{2a}(0,0)^{6h}(\uparrow \uparrow 0, \downarrow \downarrow \downarrow)$} & \textbf{-1.46} & \textbf{-0.33} & \textbf{0} \\ 
        (Ta$_{0.33}$Fe$_{0.42}$Al$_{0.25}$) & $^{2a}$(Al, Al)$^{6h}$(Fe Fe Al, Fe Fe Fe) & $^{2a}(0,0)^{6h}(\uparrow \uparrow 0, \uparrow \uparrow \uparrow)$ & 6.64 & -0.33 & -1.65 \\ 
        ~ & $^{2a}$(Al, Fe)$^{6h}$(Fe Fe Al, Fe Fe Al) & $^{2a}(0,0)^{6h}(\uparrow \uparrow 0, \downarrow \downarrow 0)$ & 0.00 & -0.33 & -2.42 \\ 
        ~ & $^{2a}$(Al, Fe)$^{6h}$(Fe Fe Al, Fe Fe Al) & \textbf{$^{2a}(0,\uparrow) ^{6h} (\uparrow \uparrow 0, \uparrow \uparrow 0)$} & 5.92 & -0.33 & -2.58 \\ 
        ~ & $^{2a}$(Al, Fe)$^{6h}$(Fe Fe Al, Al Fe Fe) & $^{2a}(0,\uparrow)^{6h}(\uparrow \uparrow 0,0 \uparrow \uparrow)$ & 6.83 & -0.33 & -2.64 \\ 
        ~ & $^{2a}$(Al, Fe)$^{6h}$(Fe Fe Al, Al Fe Fe) & $^{2a}(0,\uparrow) ^{6h} (\uparrow \uparrow 0, 0 \downarrow \downarrow)$ & 1.28 & -0.32 & -5.29 \\ 
        ~ & $^{2a}$(Fe, Fe)$^{6h}$(Fe Fe Al, Al Fe Al) & $^{2a}(\uparrow, \uparrow) ^{6h} (\uparrow \uparrow 0, 0 \uparrow 0)$ & 6.12 & -0.31 & -6.07 \\ 
        ~ & $^{2a}$(Fe, Fe)$^{6h}$(Fe Fe Al, Fe Al Al) & $^{2a}(\uparrow, \uparrow)^{6h}(\uparrow \uparrow 0, \downarrow 0 0)$ & 2.50 & -0.31 & -6.63 \\ 
        ~ & $^{2a}$(Al, Fe)$^{6h}$(Fe Al Al, Fe Fe Fe) & $^{2a}(0,\uparrow)^{6h} (\uparrow 0 0, \uparrow \uparrow \uparrow)$ & 6.45 & -0.30 & -10.98 \\ 
        ~ & $^{2a}$(Al, Fe)$^{6h}$(Fe Al Al, Fe Fe Fe) & $^{2a} (0,\uparrow) ^{6h} ( \downarrow 0 0, \uparrow \uparrow \uparrow)$ & 3.41 & -0.30 & -11.45 \\ 
        ~ & $^{2a}$(Al, Fe)$^{6h}$(Fe Al Al, Fe Fe Fe) & {$^{2a}(0,\uparrow)^{6h} (\uparrow 0 0, \downarrow \downarrow \downarrow)$} & -1.46 & -0.29 & -11.89 \\ 
        ~ & $^{2a}$(Fe, Fe)$^{6h}$(Fe Fe Al, Al Fe Al) & $^{2a}(\uparrow, \uparrow) ^{6h} (\downarrow \downarrow 0, 0 \uparrow 0)$ & 0.27 & -0.29 & -13.15 \\ 
        ~ & $^{2a}$(Fe, Fe)$^{6h}$(Al Al Al, Fe Fe Fe) & $^{2a} (0,0) ^{6h} (0 0 0, \uparrow \uparrow \uparrow)$ & 2.94 & -0.24 & -28.36 \\ \hline
        Ta$_4$Fe$_4$Al$_4$ & \textbf{$^{2a}$(Al, Al)$^{6h}$(Fe Fe Al, Fe Fe Al)} & \textbf{$^{2a}(0,0)^{6h}(\uparrow \uparrow 0, \downarrow \downarrow 0)$} & \textbf{0.00} & \textbf{-0.34} & \textbf{0} \\ 
        (Ta$_{0.33}$Fe$_{0.33}$Al$_{0.33}$) & $^{2a}$(Al, Al)$^{6h}$(Fe Fe Al, Fe Fe Al) & $^{2a}(0, 0)^{6h} (\uparrow \uparrow 0, \uparrow \uparrow 0)$ & 5.44 & -0.34 & -1.53 \\ 
        ~ & $^{2a}$(Al, Al)$^{6h}$(Fe Fe Al, Al Fe Fe) & $^{2a} (0, 0) ^{6h} (\uparrow \uparrow 0, 0 \downarrow \downarrow)$ & 0.00 & -0.34 & -1.56 \\ 
        ~ & $^{2a}$(Al, Al)$^{6h}$(Fe Fe Al, Al Fe Fe) & $^{2a} (0, 0)^{6h} (\uparrow \uparrow 0, 0 \uparrow \uparrow)$ & 5.40 & -0.34 & -1.59 \\ 
        ~ & $^{2a}$(Al, Fe)$^{6h}$(Fe Fe Al, Fe Al Al) & $^{2a} (0, \uparrow) ^{6h} (\downarrow \downarrow 0, \uparrow 0 0)$ & 1.88 & -0.34 & -2.14 \\ 
        ~ & $^{2a}$(Al, Fe)$^{6h}$(Fe Fe Al, Fe Al Al) & $^{2a}(0, \downarrow )^{6h} (\downarrow \downarrow 0 , \uparrow 0 0)$ & -1.87 & -0.34 & -2.14 \\ 
        ~ & $^{2a}$(Al, Al)$^{6h}$(Fe Al Al, Fe Fe Fe) & $^{2a} (0, 0) ^{6h} (\uparrow 0 0, \downarrow \downarrow \downarrow)$ & -3.00 & -0.34 & -2.26 \\ 
        ~ & $^{2a}$(Al, Fe)$^{6h}$(Fe Fe Al, Fe Al Al) & $^{2a} (0, \uparrow) ^{6h} (\uparrow \uparrow 0, \uparrow 0 0)$ & 4.84 & -0.33 & -3.01 \\ 
        ~ & $^{2a}$(Al, Fe)$^{6h}$(Fe Fe Al, Al Fe Al) & $^{2a} (0, \uparrow) ^{6h} (\uparrow \uparrow 0 , 0 \uparrow 0)$ & 5.48 & -0.33 & -4.41 \\ 
        ~ & $^{2a}$(Al, Al)$^{6h}$(Fe Al Al, Fe Fe Fe) & $^{2a} (0, 0) ^{6h} (\uparrow 0 0, \uparrow \uparrow \uparrow)$ & 4.83 & -0.33 & -4.67 \\ 
        ~ & $^{2a}$(Fe, Fe)$^{6h}$(Fe Al Al, Fe Al Al) & $^{2a}(\downarrow, \uparrow) ^{6h} (\downarrow 0 0 , \uparrow 0 0)$ & 0.00 & -0.33 & -4.81 \\ 
        ~ & $^{2a}$(Fe, Fe)$^{6h}$(Fe Al Al, Fe Al Al) & $^{2a} (\uparrow, \uparrow) ^{6h} (\uparrow 0 0 , \downarrow 0 0)$ & 1.13 & -0.33 & -4.99 \\ 
        ~ & $^{2a}$(Al, Fe)$^{6h}$(Fe Fe Al, Al Fe Al) & $^{2a} (0, \uparrow )^{6h} (\uparrow \uparrow 0 , 0 \downarrow 0)$ & 2.32 & -0.33 & -5.19 \\ 
        ~ & $^{2a}$(Fe, Fe)$^{6h}$(Al Fe Al, Al Fe Al) & $^{2a} (\uparrow, \uparrow )^{6h} ( 0 \uparrow 0, 0 \uparrow 0)$ & 5.45 & -0.33 & -5.74 \\ 
        ~ & $^{2a}$(Fe, Fe)$^{6h}$(Fe Al Al, Fe Al Al) & $^{2a} (\uparrow, \uparrow )^{6h} (\uparrow 0 0, \uparrow 0 0)$ & 2.89 & -0.32 & -5.85 \\ 
        ~ & $^{2a}$(Al, Fe)$^{6h}$(Fe Fe Al, Al Fe Al) & $^{2a} (0, \uparrow)^{6h} (\downarrow \downarrow 0, 0 \uparrow 0)$ & -0.79 & -0.32 & -6.00 \\ 
        ~ & $^{2a}$(Fe, Fe)$^{6h}$(Fe Al Al, Fe Al Al) & $^{2a}(\downarrow, \downarrow ) ^{6h} (\uparrow 0 0 , \uparrow 0 0)$ & 0.15 & -0.32 & -6.54 \\ 
        ~ & $^{2a}$(Fe, Fe)$^{6h}$(Al Fe Al, Al Fe Al) & $^{2a} (\uparrow, \downarrow)^{6h} (0 \uparrow 0, 0 \downarrow    0)$ & 0.00 & -0.32 & -7.56 \\ 
        ~ & $^{2a}$(Fe, Fe)$^{6h}$(Al Fe Al, Al Fe Al) & $^{2a}( \uparrow, \uparrow) ^{6h} (0 \downarrow 0, 0 \downarrow 0)$ & 0.00 & -0.31 & -11.09 \\ 
        ~ & $^{2a}$(Fe, Fe)$^{6h}$(Fe Fe Al, Al Al Al) & $^{2a}( \uparrow, \uparrow) ^{6h} (\uparrow \uparrow 0 , 0 0 0 )$ & 5.18 & -0.30 & -13.25 \\ 
        ~ & $^{2a}$(Al, Fe)$^{6h}$(Al Al Al, Fe Fe Fe) & $^{2a}(0, \uparrow) ^{6h} (0 0 0 , \uparrow \uparrow \uparrow)$ & 5.09 & -0.26 & -23.47 \\ \hline
        Ta$_4$Fe$_3$Al$_5$ & \textbf{$^{2a}$(Al, Al)$^{6h}$(Fe Fe Al, Fe Al Al)} & \textbf{$^{2a} (0, 0) ^{6h} (\downarrow \downarrow 0, \uparrow 0 0)$} & \textbf{-1.43} & \textbf{-0.36} & \textbf{0} \\ 
        (Ta$_{0.33}$Fe$_{0.25}$Al$_{0.42}$) & $^{2a}$(Al, Al)$^{6h}$(Fe Fe Al, Fe Al Al) & $^{2a} (0, 0) ^{6h} (\uparrow \uparrow 0, \uparrow 0 0)$ & 4.26 & -0.35 & -0.85 \\ 
        ~ & $^{2a}$(Al, Al)$^{6h}$(Fe Fe Al, Al Fe Al) & $^{2a} (0, 0) ^{6h} (\uparrow \uparrow 0 , 0 \downarrow 0)$ & 1.27 & -0.35 & -2.60 \\ 
        ~ & $^{2a}$(Al, Fe)$^{6h}$(Fe Al Al, Fe Al Al) & $^{2a} (0, \downarrow ) ^{6h} (\downarrow 0 0, \downarrow 0 0)$ & -4.40 & -0.35 & -2.65 \\ 
        ~ & $^{2a}$(Al, Al)$^{6h}$(Fe Fe Al, Al Fe Al) & $^{2a} (0, 0) ^{6h} (\uparrow \uparrow 0, 0 \uparrow 0)$ & 4.56 & -0.34 & -2.93 \\ 
        ~ & $^{2a}$(Al, Fe)$^{6h}$(Al Fe Al, Al Fe Al) & $^{2a}(0, \downarrow) ^{6h} (0 \downarrow 0, 0 \downarrow 0)$ & -4.93 & -0.34 & -3.06 \\ 
        ~ & $^{2a}$(Al, Fe)$^{6h}$(Fe Al Al, Fe Al Al) & $^{2a}(0, \downarrow)^{6h}(\uparrow 0 0 , \downarrow 0 0 )$ & -0.58 & -0.34 & -3.37 \\ 
        ~ & $^{2a}$(Al, Fe)$^{6h}$(Al Fe Al, Al Fe Al) & $^{2a}(0, \uparrow )^{6h}(0 \uparrow 0, 0 \downarrow 0)$ & 1.54 & -0.34 & -5.05 \\ 
        ~ & $^{2a}$(Fe, Fe)$^{6h}$(Fe Al Al, Al Al Al) & $^{2a} (0,0)^{6h} (\downarrow 0 0 , 0 0 0 )$ & -1.60 & -0.32 & -9.10 \\ 
        ~ & $^{2a}$(Fe, Fe)$^{6h}$(Fe Al Al, Al Al Al) & $^{2a}(\uparrow, \downarrow) ^{6h} (0 0 0 , 0 0 0 )$ & 0.00 & -0.32 & -9.74 \\ 
        ~ & $^{2a}$(Al, Fe)$^{6h}$(Al Fe Al, Al Fe Al) & $^{2a}(0,0)^{6h} (0 0 0, 0 0 0)$ & 0.00 & -0.32 & -10.29 \\ 
        ~ & $^{2a}$(Fe, Fe)$^{6h}$(Fe Al Al, Al Al Al) & $^{2a}(\uparrow, \uparrow) ^{6h} (0 0 0 , 0 0 0 )$ & 0.80 & -0.32 & -10.84 \\ 
        ~ & $^{2a}$(Al, Fe)$^{6h}$(Fe Fe Al, Al Al Al) & $^{2a}(0, \uparrow)^{6h}(\uparrow \uparrow 0, 0 0 0 )$ & 4.82 & -0.31 & -13.7 \\ 
        ~ & $^{2a}$(Al, Al)$^{6h}$(Al Al Al, Fe Fe Fe) & $^{2a}(0,0)^{6h}( 0 0 0 ,\uparrow \uparrow \uparrow)$ & 5.32 & -0.28 & -22.23 \\ \hline
        Fe$_2$Al$_6$Ta$_4$ & \textbf{$^{2a}$(Al, Al)$^{6h}$(Fe Al Al ,Fe Al Al)} & \textbf{$^{2a}(0,0) ^{6h}(\downarrow 00,\uparrow 00 )$} & \textbf{0.00} & \textbf{-0.36} & \textbf{0} \\ 
        (Ta$_{0.33}$Fe$_{0.17}$Al$_{0.5}$) & $^{2a}$(Al, Al)$^{6h}$(Fe Al Al ,Fe Al Al) & $^{2a}(0,0) ^{6h}(\uparrow 00, \uparrow 00 )$ & 3.58 & -0.36 & -0.21 \\ 
        ~ & $^{2a}$(Al, Al)$^{6h}$(Al Fe Al, Al Fe Al) & $^{2a}(0, 0 )^{6h} ( 0 \uparrow 0 , 0 \uparrow 0)$ & 3.73 & -0.35 & -2.99 \\ 
        ~ & $^{2a}$(Fe, Fe)$^{6h}$(Al Al Al, Al Al Al) & $^{2a}(\uparrow, \downarrow)^{6h}(000,000)$ & 0.00 & -0.35 & -3.09 \\ 
        ~ & $^{2a}$(Fe, Fe)$^{6h}$(Al Al Al, Al Al Al) & $^{2a}(\uparrow , \uparrow) ^{6h} (000,000)$ & 2.98 & -0.35 & -3.87 \\ 
        ~ & $^{2a}$(Al, Al)$^{6h}$(Al Fe Al, Al Fe Al) & $^{2a}(0,0)^{6h}(0 \downarrow 0,0 \uparrow 0)$ & 0.00 & -0.34 & -4.65 \\ 
        ~ & $^{2a}$(Al, Fe)$^{6h}$(Fe Al Al, Al Al Al) & $^{2a}(0,\downarrow)^{6h}(000,0 0 0)$ & -0.73 & -0.32 & -12.63 \\ 
        ~ & $^{2a}$(Al, Fe)$^{6h}$(Fe Al Al, Al Al Al) & $^{2a}(0,0)^{6h}(000,000)$ & -0.01 & -0.31 & -12.91 \\ 
        ~ & $^{2a}$(Al, Al)$^{6h}$(Fe Fe Al, Al Al Al) & $^{2a}(0,0 )^{6h} (\uparrow \uparrow 0,000)$ & 3.20 & -0.30 & -16.07 \\ \hline

\caption{The list of all unique Ta(Fe$_{1-x}$Al$_x$)$_2$ site occupation decorations for 7 compositions in $0 \leq x \leq 0.50$. The Wyckoff positions are denoted as superscripts. Commas separate atoms of different atomic layers along the $\langle c \rangle$ axis. The sites for the atomic and magnetic configurations are arranged in the same order as indexed in Figure \ref{fig:1}. The total magnetic moment is given per conventional unit cell. The formation energy E$^{form}$ with respect to bulk Fe, Al and Ta is given per atom with the units of eV/atom. The energy difference in percentage to the lowest formation energy is given by $\Delta$ E$_{min}^{form}$. The lowest-energy configurations are highlighted in bold.}
\label{tab:1}
\end{longtable}

\begin{table}[!ht]
    \centering
    \begin{tabular}{|l|l|}
    \hline
        Structure &  $\Delta$E$_{hull}$, eV/atom \\ \hline
        TaAl$_2$ &  0.10 \\ 
        Ta$_4$Al$_3$Fe$_5$  & 0.10 \\ 
        TaAlFe & 0.07 \\ 
        Ta$_2$AlFe$_3$ &  0.06 \\ 
        Al$_8$Fe$_5$ &  0.05 \\ 
        Ta$_4$AlFe$_7$ &0.03 \\ 
        Ta$_4$Al$_5$Fe$_3$ & 0.03 \\ 
        Ta$_{24}$Al$_{19}$ &  0.02 \\ \hline

    \end{tabular}
    \caption{The energy differences of metastable phases in the Ta-Fe-Al phase diagram relative to the convex hull.}
    \label{tab:2}
\end{table}

\twocolumn 

\bibliographystyle{elsarticle-num} 
\bibliography{cite}

\end{document}